\documentclass[12pt]{article}

\usepackage{amsfonts}
\usepackage{amsmath}
\usepackage{esint}
\usepackage[margin=1in]{geometry}
\usepackage{graphicx}
\usepackage{epstopdf}
\usepackage{amssymb}
\usepackage{subcaption}
\usepackage{float}
\usepackage{caption}
\usepackage{color}
\usepackage{array}
\usepackage{url}
\usepackage{xcolor}
\usepackage{multicol}
\usepackage{hyperref}
\usepackage{wrapfig}

\usepackage[round, semicolon, authoryear, sort]{natbib}

\definecolor{myorange}{rgb}{0.8500 0.3250 0.0980}
\definecolor{myyellow}{rgb}{0.9290 0.6940 0.1250}
\definecolor{mygreen}{rgb}{0.4660 0.6740 0.1880}
\definecolor{mypurple}{rgb}{0.4940 0.1840 0.5560}
\definecolor{mybrown}{rgb}{0.53 0.26 0.12}
\definecolor{myred}{rgb}{1 0 0}
\definecolor{mycyan}{rgb}{0 1 1}

\numberwithin{equation}{section}

\newcommand{\ds} {\displaystyle}
\newcommand{\Frac}[2]{\ds \frac{#1}{#2}}

\newcommand{\R}{{\mathbb R}}

\title{\Large A data-informed mathematical model of microglial cell dynamics during ischemic stroke in the middle cerebral artery} 
\author{\large Sara Amato$^{1}$, Andrea Arnold$^{1,2,*}$}
\date{}

\begin{document}
\maketitle

\small 

\centerline{$^1$ Bioinformatics \& Computational Biology Program, Worcester Polytechnic Institute, Worcester, MA, USA}
\vspace{.1cm}

\centerline{$^2$ Department of Mathematical Sciences, Worcester Polytechnic Institute, Worcester, MA, USA}
\vspace{.2cm}

\centerline{* Corresponding author: anarnold@wpi.edu}

\bigskip

\begin{abstract}
Neuroinflammation immediately follows the onset of ischemic stroke in the middle cerebral artery. During this process, microglial cells are activated in and recruited to the penumbra. Microglial cells can be activated into two different phenotypes: M1, which can worsen brain injury; or M2, which can aid in long-term recovery. In this study, we contribute a summary of experimental data on microglial cell counts in the penumbra following ischemic stroke induced by middle cerebral artery occlusion (MCAO) in mice and compile available data sets into a single set suitable for time series analysis. Further, we formulate a mathematical model of microglial cells in the penumbra during ischemic stroke due to MCAO. Through use of global sensitivity analysis and Markov Chain Monte Carlo (MCMC)-based parameter estimation, we analyze the effects of the model parameters on the number of M1 and M2 cells in the penumbra and fit identifiable parameters to the compiled experimental data set. We utilize results from MCMC parameter estimation to ascertain uncertainty bounds and forward predictions for the number of M1 and M2 microglial cells over time. Results demonstrate the significance of parameters related to M1 and M2 activation on the number of M1 and M2 microglial cells. Simulations further suggest that potential outliers in the observed data may be omitted and forecast predictions suggest a lingering inflammatory response. \\

\end{abstract}

\normalsize

\section{Introduction}

The aim of this study is to develop a data-informed mathematical model of microglial cell dynamics in the penumbra following ischemic stroke. 
Stroke is a leading cause of death and disability in the United States \citep{Johnson2016, Kim2020}. Ischemic strokes, which account for 87\% of all strokes, are caused by a blocked blood vessel in the brain \citep{ASA}. In this work, we focus on ischemic stroke due to obstruction of the middle cerebral artery \citep{Moulin1985, Feske2021, Uzdensky2019}, referred to as middle cerebral artery occlusion (MCAO).
Following ischemic stroke due to MCAO, two main affected areas of the brain emerge: the ischemic core, which is characterized by irreversible damage; and the penumbra, which can be salvaged with timely intervention and is the most clinically relevant target for treatment \citep{Feske2021, Uzdensky2019}.

Neuroinflammation immediately follows the onset of ischemic stroke and begins with the activation and recruitment of resting microglial cells in the penumbra \citep{Feske2021, Uzdensky2019}. Microglia activation from the resting state is characterized by two phenotypes: M1 cells (classical activation), which can exacerbate the inflammatory response and lead to further brain damage; and M2 cells (alternative activation), which can limit inflammation \citep{Hu2015, Tang2016, Hao2016, Orihuela2016}.
During the neuroinflammatory process, M1 and M2 microglial cells undergo dynamic interactions with each other. In particular, once activated, microglial cells may switch phenotypes from M1 to M2 and vice versa \citep{Zhao2017, Hu2015, Tanaka2015, Orihuela2016, Qin2019, Nakagawa2014, Guruswamy2017}; however, the switching from the M2 to M1 phenotype has been cited as an area in need of further research \citep{Boche2013}.

In this work, we contribute a new mathematical model focusing on the microglial response to MCAO-induced ischemic stroke which includes bidirectional phenotype switching in order to accommodate these potentially influential interactions. 
Previous mathematical models of neuroinflammation have included M1 and M2 microglial phenotypes for applications other than stroke, including traumatic brain injury (TBI), amyotrophic lateral sclerosis
(ALS), hemorrhagic shock, and Alzheimer’s disease \citep{Vaughan2018, Shao2013, Hao2016}. The models for TBI in \cite{Vaughan2018} and for Alzheimer’s disease in \cite{Hao2016} include switching from M1 to M2 but do not include switching from the M2 to the M1 phenotype. The model for ALS presented in \cite{Shao2013} includes bidirectional switching between microglia phenotypes. Further studies have explored the interactions between cytokines in general neuroinflammation \citep{Anderson2015, Torres2009} but have not included the interactions of microglia producing these substances. Several studies have also explored inflammation with macrophages, which behave in a similar manner to microglia, in applications such as myocardial infarction \citep{Malek2015, Wang2012}. Other papers have considered microglial cells and ischemic stroke but have not used parameter estimation techniques to fit the models to experimental data \citep{Dumont2013, Chapuisat2008, DiRusso2010, Alqarni2021}.

\subsection{Paper Contributions}

In a previous study, we contributed a system of nonlinear differential equations modeling the dynamics of interacting microglia and cytokines in the general ischemic brain \citep{Amato2021}. We modeled the dynamics over a period of three days and set nominal parameter values to uphold trends found in the literature. Although this model provides a starting point for analyzing key neuroinflammatory components during ischemic stroke, it is limited due to its short time duration, premature shift in M2 to M1 dominance, assumption that the whole ischemic brain behaves in a similar manner, and lack of fit to experimental data. Motivated by these limitations, in this work we construct a new data-informed mathematical model where we fit the new model parameters using a collection of data from experimental studies.
The main contributions of this work include the following:
\begin{itemize}
\item We provide a summary of experimental data on microglial cells in the penumbra post MCAO-induced ischemic stroke in mice models and compile data from similar studies into a single set of time series observations suitable for parameter estimation.
\item We formulate a new data-informed mathematical model of microglial cells in the penumbra during MCAO-induced ischemic stroke and use this model to analyze and predict the number of microglial cells over time.
\end{itemize}
In developing this model, we outline a procedure that combines mathematical modeling with computational techniques from sensitivity analysis, parameter estimation, and uncertainty quantification to make data-informed predictions of microglial cell counts throughout the neuroinflammatory process.

Most experimental studies in the literature collect data on microglial cells during MCAO-induced ischemic stroke at only a few time points over a fourteen day period. To accurately formulate a data-informed mathematical model and fit the model parameters using time series data, we need a data set with measurements of the cell counts at multiple time points throughout the stroke duration. In this study, we fill this gap by compiling data from a number of comparable studies using permanent and transient MCAO mice models. We combine data from multiple studies to provide one cell count (given in number of microglial cells per millimeter squared) for M1 and M2 cells in the penumbra during MCAO-induced ischemic stroke at zero, one, two, three, five, seven, and fourteen days following stroke onset. This contribution has the potential to further knowledge of microglial cell dynamics by providing data in a format that can be utilized in future computational studies.

Further, we develop a new mathematical model using coupled nonlinear differential equations to describe the microglial cell dynamics in the penumbra during MCAO-induced ischemic stroke. To the authors' knowledge, this paper presents the first mathematical model to focus specifically on these dynamics. Moreover, while previous mathematical models of ischemic stroke and neuroinflammation have set parameter values based on trends and biological assumptions, many have not fit these terms to experimental data. 
In this paper, we carry out a robust estimation procedure to fit the most sensitive parameters in the proposed model to the compiled experimental data set and use the data-informed model to make forecast predictions.
We make use of a Monte Carlo sampling approach to not only estimate these model parameters based on the observed data but also provide a measure of uncertainty in the estimated values, encoded in posterior probability distributions.
This further allows for forward propagation of uncertainty in the model predictions.

\subsection{Paper Organization}

The paper is organized as follows. Section \ref{sec: Data} describes a compiled set of experimental data on microglial cell counts during MCAO-induced ischemic stroke in mice. Section \ref{sec: Model} derives a new coupled system of nonlinear differential equations with nominal parameter values chosen to model the trends of M1 and M2 microglial cells in the penumbra post MCAO-induced ischemic stroke observed in the experimental data. Section \ref{sec: robustest} details a robust estimation procedure to fit model parameters to the compiled set of experimental data and provides forecast predictions of M1 and M2 microglial cells with forward propagation of uncertainty.  
Section \ref{sec: Discussion} features a discussion of the results, conclusions, and future work.

\section{Experimental Data on Microglial Cell Counts} \label{sec: Data} 

In this section, we summarize eight experimental studies that include data on the number of M1 and M2 microglial cells in the penumbra post MCAO-induced ischemic stroke in mice. Motivated by the goal of contributing a new data-inspired mathematical model of microglial cells in the neuroinflammatory process after ischemic stroke, we compile data from the presented experimental studies into a single set of time series data to use within our modeling procedure.

Experimental data reporting M1 and M2 cell measurements for this type of stroke model is limited in the literature, and most of the available studies provide infrequent time measurements of the microglial cell counts.
For example, four of the studies considered in this work provide the number of cells at only one time point, and two studies provide measurements at two time points. 
Data sets with infrequent measurements limit our ability to reliably estimate the parameters of a dynamic model over the time duration of interest (here over a period of 14 days).
While the remaining two studies provide measurements of the microglial cell counts more frequently, i.e., at five and six time points, respectively, reliance on these data sets alone could bias the estimated parameters (e.g., due to measurement errors and possible outliers in the data).
Therefore, we aim to combine the available data sets into a single set of time series data that accounts for the observations across different studies and can be used to estimate parameters of a dynamical system.

Transient and permanent MCAO are commonly used biological stroke models in both mice and rats. During these procedures, researchers insert a suture to the middle cerebral artery of a rodent to block blood flow, resulting in ischemic stroke. In the transient case, researchers remove the suture from the middle cerebral artery to restore blood flow, whereas, in the permanent case, researchers do not remove the suture from the middle cerebral artery. Both are accepted as models that closely simulate human ischemic stroke \citep{Fluri2015}. Due to the prevalence of this procedure, only experimental studies using the MCAO with suture method are included in the compiled data for this study.

In all eight studies considered, researchers induce ischemic stroke in adult male mice by blocking the left middle cerebral artery with a silicon suture. The occlusion remains for a period of 40 minutes on, after which researchers either remove the suture or keep it in for the duration of the experiment. If MCAO is transient, measurements taken at Day 1 are taken one day after blood flow is restored. If MCAO is permanent, Day 1 is considered to be one day after the procedure takes place. At each day that measurements are taken, multiple mice are sacrificed and their brains are used by researchers to collect data on the number of M1 and M2 cells. Note that the number of mice sacrificed at each time point remains consistent within each study but varies across the different studies. For example, in the study of \cite{Li2018}, three mice are sacrificed at Day 1 and three mice are sacrificed at Day 3, but in \cite{Ma2020}, five mice are sacrificed at Day 2.

Once the mice are sacrificed, researchers collect the brains of the mice and image a cross-section of the penumbra, using ImageJ or fluorescence confocal microscopy to count automatically recognized cells.  To recognize cells, they perform immunohistochemistry, which is a process that uses antibodies to detect antigens within the brain regions. M1 and M2 phenotypes express distinct antibodies, including CD16 and CD32 for M1 and CD206 and Iba1 for M2. All studies considered in this work use one or both of these antibodies to identify the respective phenotype. M1 and M2 cell counts are taken at various times and are presented as an average over the number of mice and number of cross-sections imaged and are measured in cells per millimeter squared. Table \ref{table: papers} provides a description and reference of the paper for each study, as well as summarizes the differences across studies. Note that \cite{Suenaga2015} is the only experimental study which utilizes the permanent MCAO model; we denote this distinction in Table \ref{table: papers} by using a `$\infty$' symbol in the occlusion time column.

\begin{table}[t!]
\centering
\renewcommand{\arraystretch}{1.2}
\begin{tabular}{| p{.25\linewidth} | p{.1\linewidth} | p{.2\linewidth} | p{.15\linewidth} | p{.18\linewidth} |} 
 \hline
Reference & $\#$ Mice & M1 $\&$ M2 Markers & Occlusion Time (Mins) & Measurement Times (Days)\\
 \hline\hline
\cite{Hu2012} & 6 & CD16/CD32, CD206 & 40 & 0, 1, 3, 5, 7, 14\\
    \hline
 \cite{Li2018} & 3 & CD16, CD206 & 45 & 1, 3\\
    \hline
 \cite{Ma2020} & 5 & CD16/CD32, CD206/Iba1  & 60 & 2\\
    \hline
\cite{Wang2017} & 9 & CD16, CD206 & 45 & 14\\
    \hline
\cite{Xu2021} & 6  & CD16/CD32, CD206 & 60 & 3\\
    \hline
 \cite{Li2023} & 5 & CD16/CD32, CD206 & 90 & 3\\
    \hline
\cite{Suenaga2015} & 4 & CD16/CD32, CD206 & $\infty$ & 0, 1, 3, 7, 14\\
    \hline
 \cite{Yang2017} & 3-5 & CD16, CD206 & 60 & 3, 7\\
\hline 
\end{tabular}
\caption{Reference and summary of experimental studies used in the compiled data set. The name and reference of each paper is given, as well as the number of mice used per measurement group, the M1 and M2 cell markers utilized to identify the type of cell in the counting procedure, the length of the occlusion time in minutes, and the days that the measurements are taken. Note that the `$\infty$' symbol means that permanent MCAO was used.}
\label{table: papers}
\end{table}

\begin{figure}[t!]
\centerline{\includegraphics[width=1\linewidth]{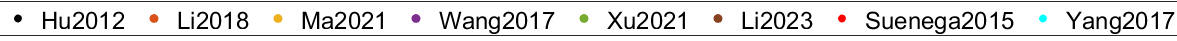}}
\vspace{0.1cm}
\centerline{\includegraphics[width=.5\linewidth]{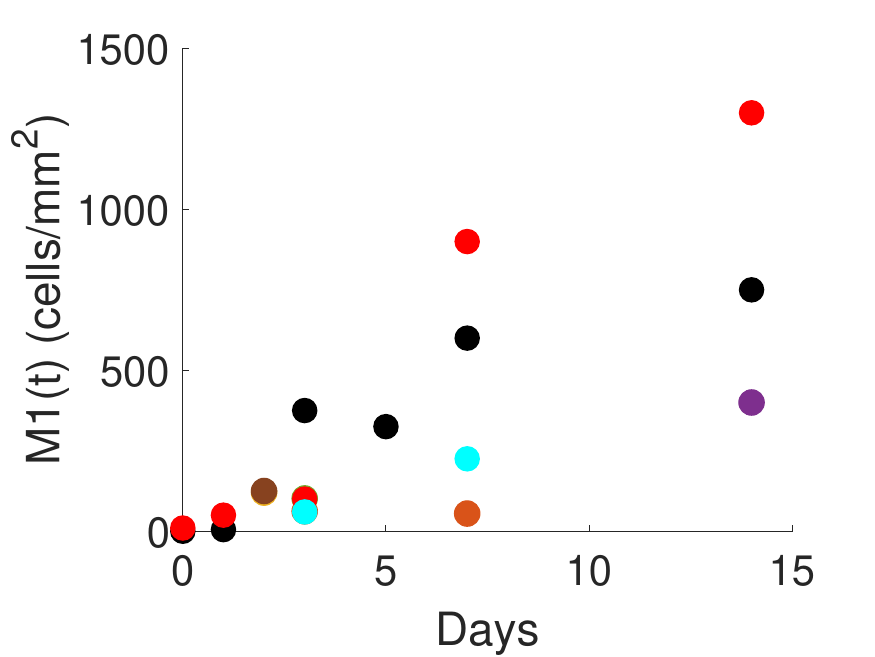} \includegraphics[width=.5\linewidth]{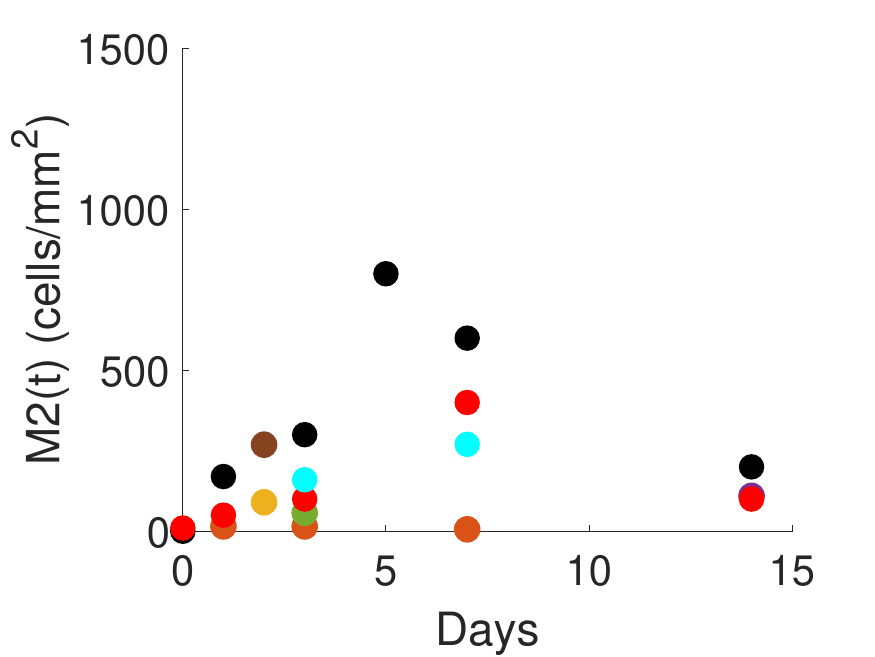}}
\vspace{0.4cm}
\begin{minipage}{0.5 \textwidth} 
\centering
\begin{tabular}{|c c c c c c c|} 
 \hline
 0 & 1 & 2 & 3 & 5 & 7 & 14\\  
 \hline\hline
 0 & 5 & & 375 & 325 & 600 & 750 \\
    \hline
 &  & & \textcolor{myorange}{62} & & \textcolor{myorange}{55}  & \\
    \hline
 & & \textcolor{myyellow}{120} & & & & \\
    \hline
& & & & & & \textcolor{mypurple}{400} \\
    \hline
& & &  \textcolor{mygreen}{102} & & & \\
    \hline
 & & \textcolor{mybrown}{125} & &  & &  \\
    \hline
 \textcolor{myred}{10} & \textcolor{myred}{50} & & \textcolor{myred}{100} & &  \textcolor{myred}{900} & \textcolor{myred}{1300} \\
    \hline
 & & & \textcolor{mycyan}{60} & & \textcolor{mycyan}{225} & \\
\hline 
\end{tabular}
    \end{minipage}
    \hfill
    \hspace{10mm}
    \begin{minipage}{0.5 \textwidth}
\begin{tabular}{|c c c c c c c|} 
 \hline
0 & 1 & 2 & 3 & 5 & 7 & 14\\ 
 \hline\hline
 0 & 170 & & 300 & 800 & 600 & 200 \\
    \hline
 & \textcolor{myorange}{15} & & \textcolor{myorange}{15} & & \textcolor{myorange}{6}  & \\
    \hline
 & & \textcolor{myyellow}{90} & & & & \\
    \hline
 & & & & & & \textcolor{mypurple}{110} \\
    \hline
 & & &  \textcolor{mygreen}{57} & & & \\
    \hline
 & & \textcolor{mybrown}{269} & &  & &  \\
    \hline
 \textcolor{myred}{10} & \textcolor{myred}{50} & & \textcolor{myred}{100} & &  \textcolor{myred}{400} & \textcolor{myred}{100} \\
    \hline
& & & \textcolor{mycyan}{160} & & \textcolor{mycyan}{270} & \\
\hline 
\end{tabular}
\label{table: dataM2}
\end{minipage}
\caption{Compiled data from the experimental studies summarized in Table~\ref{table: papers}. Number of M1 (left column) and M2 (right column) microglial cells displayed in both plot and table formats. Colors correspond to the above legend. Note that the first row of each table lists the measurement days, and the cell counts are measured in cells per millimeter squared.}
\label{fig: rawdata}
\end{figure}

Although experimental setup is similar, counts of M1 and M2 microglial cells at the measurement times vary across studies. For M1 cells at Day 0, measurements range from 0 cells to 10 cells; at Day 1, measurements range from 5 cells to 50 cells; at Day 2 measurements range from 120 cells to 125 cells; at Day 3, measurements range from 7 cells to 375 cells; at Day 5, there is a single measurement of 325 cells; at Day 7, measurements range from 55 cells to 900 cells; and at Day 14, measurements range from 400 cells to 1300 cells. For M2 cells at Day 0, counts range from 0 cells to 10 cells; at Day 1, measurements range from 15 cells to 170 cells; at Day 2, measurements range from 90 cells to 169 cells; at Day 3, measurements range from 15 cells to 300 cells; at Day 5, there is a single measurement at 800 cells; at Day 7, measurements range from 6 cells to 600 cells; and at Day 14, measurements range from 100 cells to 200 cells. Figure \ref{fig: rawdata} shows the cell counts at each measurement time for each experimental study considered, both as a scatter plot and as a table of cell counts.

\begin{figure}[t!]
\centerline{\includegraphics[width=0.5\linewidth]{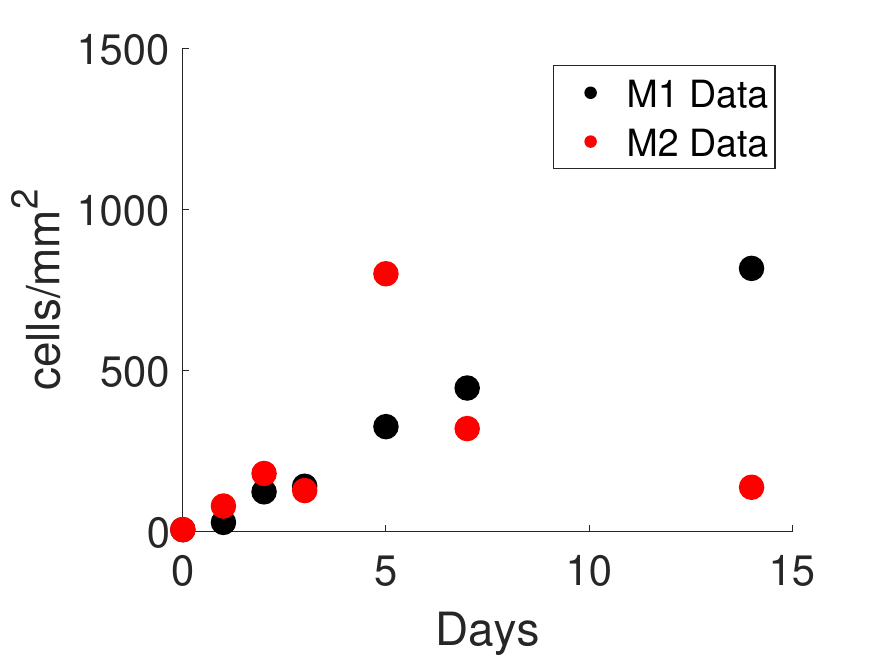}}
\vspace{0.4cm}
\centering
\begin{minipage}{0.5 \textwidth} 
\begin{tabular}{|c| c c c c c c c|} 
 \hline
Day & 0 & 1 & 2 & 3 & 5 & 7 & 14\\ 
    \hline\hline
M1 &  5 & 27.5 & 122.5 & 139.8 & 325 & 445 & 816.67 \\
    \hline
M2 &  5 & 78.33 & 179.5 & 126.4 & 800 & 319 & 136.67  \\
\hline 
\end{tabular}
\end{minipage}
\caption{Average M1 (black) and M2 (red) microglial cell measurements computed from the experimental data in Figure \ref{fig: rawdata}. The average here is calculated as the mean of the measurements of either M1 or M2 at each day. Note that the cell counts at Day 5 come from a single experimental study \citep{Hu2012}. 
} 
\label{fig: dataav}
\end{figure}

In order to obtain one time series to utilize within the modeling framework, we compute an average measurement at each time point that data are collected. At Day 5, there is only one measurement for the M1 and M2 microglial cells, so these numbers are taken as the average measurement. Taking a simple average ensures that we do not incorrectly weight measurements or put added emphasis on a point that is not indicative to the true trend of microglial cells. Note that here `average' refers to the average of the measurements over each day for the M1 and M2 cell counts given in the experimental studies. The measurements in each study are reported as averages of the number of M1 or M2 cells in the penumbra over the number of cross-sections considered and over the number of mice per measurement group. Figure \ref{fig: dataav} shows the computed average data points for the M1 and M2 microglial cells.

Note that the data at Day 5 could be considered as outliers since the number of M2 cells recorded is much higher than the measurements at Day 3 and Day 7.  Additionally, it is the only measurement at this day (collected from a single experimental study), while the cell counts for other days are averaged across multiple studies. One way to deal with this is to remove the measurements at Day 5 from the compiled data set.  Throughout the remainder of this paper, we analyze the differences in numerical results when the measurements at Day 5 are both included and omitted.


\section{Mathematical Model and Nominal Parameter Values}\label{sec: Model}

In this section, we formulate a nonlinear system of ordinary differential equations (ODEs) for modeling M1 and M2 microglial cell dynamics in the penumbra after ischemic stroke due to middle cerebral artery occlusion. We propose nominal parameter values that uphold the trends of microglial cells found in the averaged experimental data.

\subsection{Model Description}

We assume a constant source of resting microglial cells, which activates into the M1 or M2 phenotypes immediately following the onset of ischemic stroke. We assume that this activation occurs at a time-dependent rate taking into account the dominance of M2 microglial cells at the beginning of the neuroinflammatory process, followed by the eventual takeover of M1 cells \citep{Wang2017, Kigerl2009, Perego2011, Li2023, Lee2014, Ma2017}. Since there is evidence that M1 switches to the M2 phenotype and M2 switches to the M1 phenotype \citep{Hu2015, Shao2013, Tanaka2015, Vaughan2018, Nakagawa2014, Cherry2014, Boche2013, Guruswamy2017, Qin2019, Orihuela2016, Zhao2017}, we include terms for bidirectional switching within the model. We also include natural mortality for both the M1 and M2 microglia at a constant rate.

Based on these assumptions, the resulting pair of coupled ODEs describing the change in M1 and M2 microglial cells over time follows as:
\begin{eqnarray}
\frac{dM1}{dt} &=& a f_{M1}(t) - s_{M1 \rightarrow M2} M1 + s_{M2 \rightarrow M1} M2 - \mu_{M1} M1 
\label{eq: modelM1} \\[0.2cm]
\frac{dM2}{dt} &=& a f_{M2}(t) + s_{M1 \rightarrow M2} M1 - s_{M2 \rightarrow M1} M2 - \mu_{M2} M2
\label{eq: modelM2}
\end{eqnarray}
where $a$ is the number of resting microglia, $s_{M1 \rightarrow M2}$ and $s_{M2 \rightarrow M1}$ are constant rates for the microglial cell switching, $\mu_{M1}$ and $\mu_{M2}$ are constant natural mortality rates for each of the microglial cell phenotypes, and $f_{M1}(t)$ and $f_{M2}(t)$ are time-dependent functions that take into account the activation of resting microglial cells to the M1 and M2 phenotypes, respectively. 
The following sections describe the model terms in more detail.

\subsubsection{Microglial Cell Activation}

Following the onset of ischemic stroke, resting microglial cells can become polarized to the M1 phenotype or to the M2 phenotype \citep{Orihuela2016, Taylor2013, Tang2016, Nakagawa2014, Shao2013}. In the proposed model, we assume that resting microglia become activated to each phenotype at a time-varying rate, which takes into account the initial dominance of M2 microglial cells, followed by the eventual takeover of M1 cells.

This trend has been shown in experimental studies of neuroinflammation \citep{Wang2013, Kigerl2009, Perego2011, Hu2012, Li2023, Wang2017}  and has been cited in review papers \citep{Lee2014, Taylor2013, Ma2017}. Additionally, this trend is seen in traumatic brain injury \citep{Wang2013}, spinal cord injury \citep{Kigerl2009}, and in the ischemic core after ischemic stroke \citep{Perego2011}. In the proposed model, we focus on activation in the penumbra. In \cite{Hu2012}, M2 cells are dominant until Day 7 before returning to baseline and being surpassed by the growing number of M1 cells. In \cite{Li2023}, microglial cell counts are only considered at Day 2; however, at this time M2 is the dominant phenotype. In \cite{Wang2017}, microglial cell counts are only considered at Day 14; at this time, M1 is dominant. Although the trends of the two phenotypes are unclear before this two-week mark, this study may suggest that the number of M1 cells are considerably higher than the number of M2 cells two weeks post MCAO-induced ischemic stroke. Additionally, we see this trend in the averaged experimental data: M2 is the dominant phenotype until Day 5, after which the M1 cells surpass the number of M2 cells for the remainder of the time points considered. Note that if the data point at Day 5 is removed, M2 cells are still dominant at the beginning of neuroinflammation before the M1 cells takeover.

Based on these assumptions, we model the activation of resting microglial cells to the M1 or M2 phenotype using constant parameter ($a$) times time-dependent functions inspired by logistic growth and decay. In particular, we define the activation functions $f_{M1}(t)$ and $f_{M2}(t)$, which represent the rates that resting microglial cells become activated to the M1 phenotype and the M2 phenotype, respectively, as:
\begin{eqnarray}
f_{M1}(t) &=& \frac{L_{M1}}{1 + \exp{(-k_{M1}(t - t_{M1}))}} 
\label{eq: fM1} \\[0.2cm]
f_{M2}(t) &=& \frac{L_{M2}}{1 + \exp{(-k_{M2}(t - t_{M2}))}}
\label{eq: fM2}
\end{eqnarray}
where $L_{M1}$ and $L_{M2}$ are the maximum rates of activation of resting microglial cells to the M1 and M2 phenotypes, respectively, $k_{M1}$ and $k_{M2}$ are constants which define the steepness of the curves, and $t_{M1}$ and $t_{M2}$ are the midpoint values of each function.

\subsubsection{Microglial Cell Switching}

There is evidence that microglia may switch phenotypes once activated; however, the switching from the M2 to M1 phenotype has been cited as an area for further research \citep{Boche2013}. We include bidirectional switching in the proposed model, so that all potentially important dynamics driving the data are considered. In the present work, we assume that, once activated, microglia may switch from the M1 phenotype to the M2 phenotype \citep{Hu2015, Shao2013, Tanaka2015, Vaughan2018, Nakagawa2014, Cherry2014, Boche2013, Guruswamy2017, Qin2019, Orihuela2016, Zhao2017, Li2023} and from the M2 phenotype to the M1 phenotype \citep{Orihuela2016, Hu2015, Tanaka2015, Nakagawa2014, Guruswamy2017, Qin2019, Zhao2017, Li2023, Zheng2019}.

Switching from the M1 to M2 phenotype is positively influenced by the concentration of anti-inflammatory cytokines \citep{Hu2015, Shao2013}, whereas switching from the M2 to M1 phenotype is positively influenced by the concentration of pro-inflammatory cytokines \citep{Orihuela2016, Hu2015, Tanaka2015}. Experiments have been performed to ascertain the levels of different cytokines during ischemic stroke, including pro-inflammatory cytokines (IL-1$\beta$, IL-2, IL-8, and TNF-$\alpha$), as well as anti-inflammatory cytokines (IL-10 and IL-6). These studies measured the level of cytokines in human blood during acute ischemic stroke \citep{Nayak2012} and postmortem brain tissue \citep{Dziewulska2003, Tarkowski1997, Doll2014}. Results vary between studies, patients, and specific cytokines. In general, pro-inflammatory cytokines follow one of three patterns: reach a peak between 2 and 3 days after the onset of ischemic stroke before slowly returning to their original levels; remain elevated for the duration of the study; or are elevated at intermittent time periods. Anti-inflammatory cytokines additionally follow one of three patterns: peak around 2-5 days after the onset of ischemic stroke; remain elevated most of the time; or have a very small initial decrease followed by a small increase.
Since the general trends for pro- and anti-inflammatory cytokines are uncertain, we assume that the switching rates between microglial cell phenotypes remain constant throughout the time period being considered.

\subsection{Nominal Parameter Values and Simulated Output} \label{subsec: np}

Given the system of equations in \eqref{eq: modelM1}--\eqref{eq: modelM2}, we set nominal parameter values to reflect two trends found in the average data: (i) that M2 microglial cells are the dominant phenotype for the initial stages of inflammation followed by an eventual takeover by M1 microglial cells, and (ii) that M1 cells remain elevated at Day 14 while M2 cells return to baseline. Table \ref{table: nominal} gives the index, description, nominal value, and units of the parameters in equations \eqref{eq: modelM1} and \eqref{eq: modelM2} with time-dependent activation functions in equations \eqref{eq: fM1} and \eqref{eq: fM2}.

\begin{figure}[t!]
	\centering

			\begin{subfigure}{0.45\textwidth}
   		\includegraphics[width=1\linewidth]{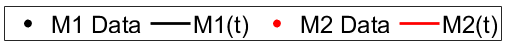}
		\includegraphics[width=1\linewidth]{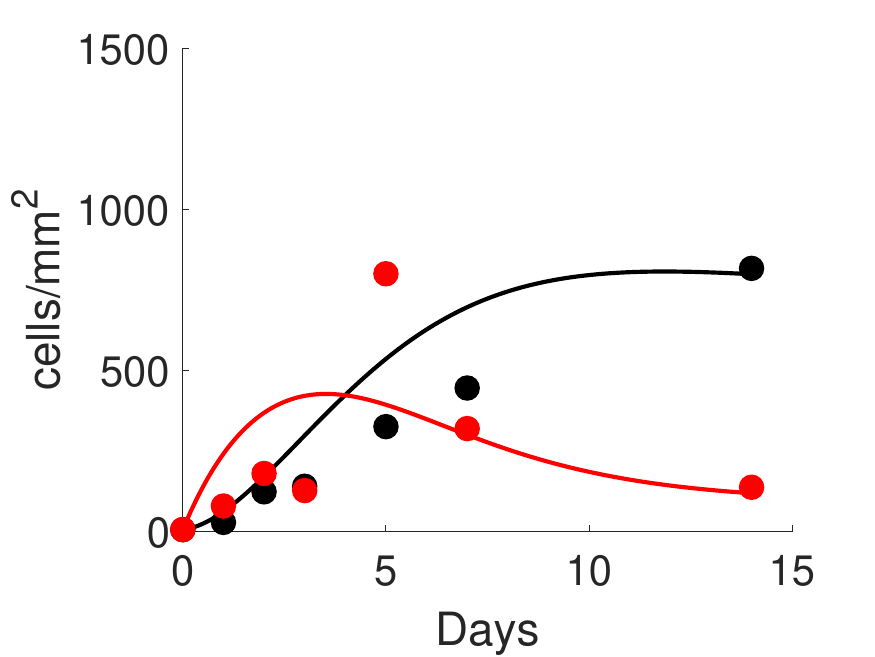}
	\end{subfigure}\hfil
	
	\medskip
		
			\begin{subfigure}{0.45\textwidth}
		\includegraphics[width=1\linewidth]{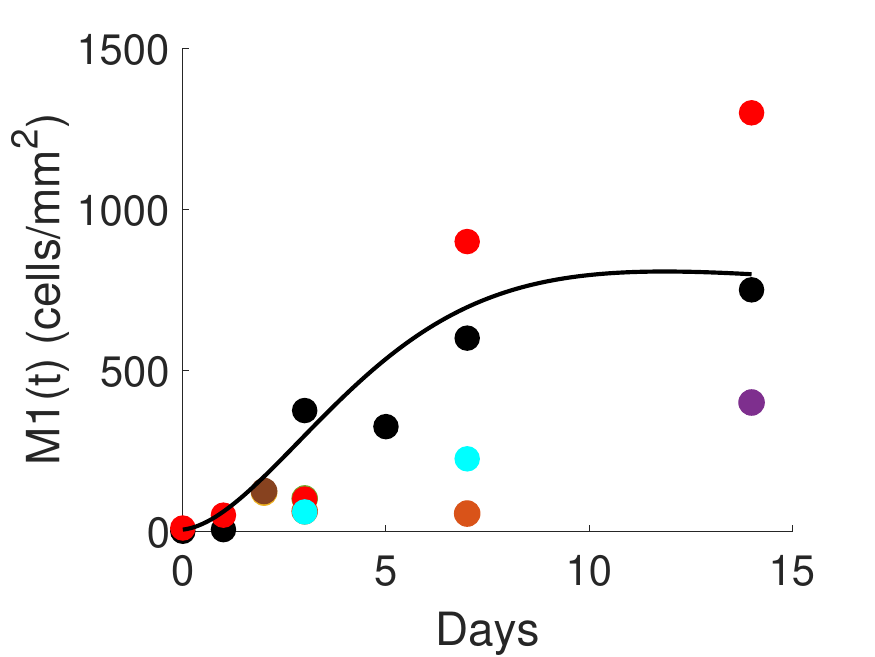}
	\end{subfigure}\hfil
			\begin{subfigure}{0.45\textwidth}
		\includegraphics[width=1\linewidth]{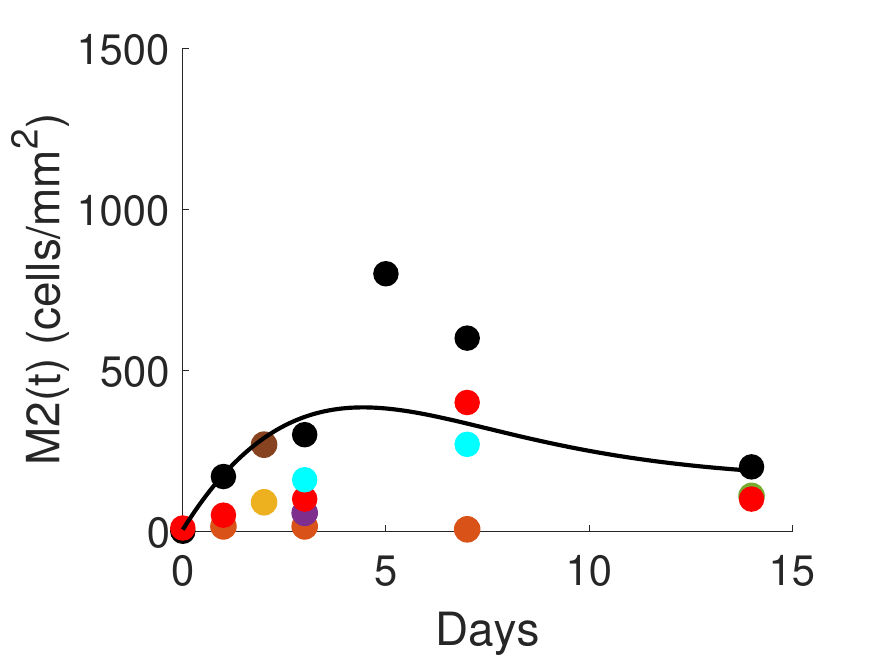}
	\end{subfigure}\hfil
	\caption{Numerical solution to model in equations \eqref{eq: modelM1} and \eqref{eq: modelM2} with time-varying activation functions in \eqref{eq: fM1} and \eqref{eq: fM2} over time interval of $0$ to $14$ days and using nominal parameter values found in Table \ref{table: nominal} and initial values set to the average data measurements at time 0, i.e, $M1(0) = M2(0) = 5$. Note that in the top figure the red and black dots indicate the averaged microglial cell data, whereas the solid lines indicate the model output. In the bottom two panels the dots correspond to the measurements from each experimental study with colors indicated in legend from Figure \ref{fig: rawdata}. Solid lines represent model outputs for M1 and M2 microglial cells respectively. }
\label{fig: modeloutputs}
\end{figure}

Figure \ref{fig: modeloutputs} shows the simulated model output using the nominal parameter values. We use MATLAB® programming language (The MathWorks, Inc., Natick, MA) and the built-in ODE solver \texttt{ode15} to compute the numerical solution to the model \eqref{eq: modelM1}--\eqref{eq: modelM2} with parameter values in Table \ref{table: nominal} and initial values set to be the average observed measurements at time 0, i.e, $M1(0) = M2(0) = 5$. Figure \ref{fig: modeloutputs} plots the simulated $M1(t)$ and $M2(t)$ curves along with the average data set (top) as well as with the full sets of experimental data for M1 and M2 microglial cells (bottom row). Note that the M2 phenotype is dominant for 3.9 days before decreasing towards its initial value, whereas after 3.9 days M1 cells keep increasing.

\begin{table}[t!]
\renewcommand{\arraystretch}{1.25}
\begin{center}
\begin{tabular}{|c c c c c|} 
 \hline
Index & Parameter & Description & Nominal Value & Unit\\
 \hline\hline
 1 & $a$ & Number of resting microglia & 500 & cells\\ 
 \hline
 2 & $L_{M1}$ & Maximum value of $f_{M1}(t)$ & 0.15 & unitless\\
 \hline
3 & $k_{M1}$ & Logistic growth rate / steepness of $f_{M1}(t)$ & 0.5 & unitless \\
 \hline
4& $t_{M1}$ & Midpoint of $f_{M1}(t)$ & 3 & days \\
 \hline
5 & $L_{M2}$ & Maximum value of $f_{M2}(t)$ & 0.75 & unitless\\
 \hline
6 & $k_{M2}$ & Logistic growth rate / steepness of $f_{M2}(t)$ &  -0.5 & unitless \\
 \hline
7& $t_{M2}$ & Midpoint of $f_{M2}(t)$ & 3 & days \\
 \hline
 8 & $s_{M1 \rightarrow M2}$ & Rate of $M1 \rightarrow M2$ switching & 0.05 & $\frac{1}{\text{days}}$ \\
 \hline
  9 & $s_{M2 \rightarrow M1}$ & Rate of $M2 \rightarrow M1$ switching & 0.325 & $\frac{1}{\text{days}}$ \\
  \hline 
  10 & $\mu_{M1}$ & Natural decay rate of M1 cells & 0.1 & $\frac{1}{\text{days}}$\\
  \hline
    11 & $\mu_{M2}$ & Natural decay rate of M2 cells & 0.1 & $\frac{1}{\text{days}}$\\ 
    \hline
\end{tabular}
\end{center}
\caption{Indices, description, and nominal values of the model parameters in equations \eqref{eq: modelM1} and \eqref{eq: modelM2} with the time-dependent activation functions given in equations \eqref{eq: fM1} and \eqref{eq: fM2}.}
\label{table: nominal}
\end{table}

\section{Robust Model Parameter Estimation and Forecast Predictions} \label{sec: robustest}

As described in the previous section, the ODE model \eqref{eq: modelM1}--\eqref{eq: modelM2} includes 11 parameters that we initially set with nominal values to uphold trends found in the experimental data. 
While the data-inspired parameters give reasonable simulation results, we did not directly incorporate the data in setting these values, and different parameter combinations may yield similar results.
In this section, we proceed with a more robust approach to incorporate the experimental data in fitting the model parameters. First, we use global sensitivity analysis techniques to identify the most influential parameters. Next, we estimate these parameters using the averaged experimental data and a Markov Chain Monte Carlo sampling approach.
Finally, once we obtain the resulting parameter distributions, we compute model forecast predictions with forward propagation of uncertainty.

\subsection{Generalized Sobol Sensitivity Analysis}

Since the behavior of the model depends on the 11 parameters summarized in Table~\ref{table: nominal} and different combinations of parameter values may yield similar results, we use global sensitivity analysis to determine which parameters are most influential to the output of M1 and M2 microglial cells. Global sensitivity approaches aim to quantify how uncertainty and variability in a single model output can be attributed to uncertainties in the input. In particular, we utilize a generalized version of the Sobol method \citep{Sobol1993}. To implement generalized Sobol sensitivity analysis in this work, we follow the algorithm described in \cite{Smith2013} with the generalized total index suggested in \cite{Gamboa2013}. In the following, we present an overview of this process for the problem at hand; see the aforementioned references for more details on the algorithms.

In Sobol sensitivity analysis, the total effect index of each parameter measures that parameter's contribution to a single output variance, including interactions with other parameters \citep{Smith2013}. Consider a scalar response variable 
\begin{eqnarray}
    Y = f(q) \label{eq: outputsobol}
\end{eqnarray}
where $q$ is a vector of model inputs. Mathematically, we define the total effect index for parameter $i$ as the complement of the variance of the expected value of $Y$ given the values of all inputs except parameter $i$, normalized by the total variance. These leads to the following expression for calculating the total effect index, denoted $S_{T_i}$, for parameter $i$:
\begin{eqnarray}
S_{T_i} = 1 - \frac{\text{var}[\mathbb{E}(Y|q_{\sim i})]}{\text{var}(Y)}. \label{eq: totalsobol}
\end{eqnarray}
In the current work, we have data for M1 and M2 microglial cells, so we would like to find the sensitivity of inputs on both model outputs. Sobol sensitivity analysis has been generalized to take into account multiple outputs \citep{Baccini2021, Gamboa2013, Gamboa2014}. This amounts to taking the weighted average of the single-output total effect indices, with weights proportional to each individual output variance. Letting $Y_1$ and $Y_2$ be the model outputs of interest, we compute the generalized total effect index of parameter $i$ as 
\begin{eqnarray}
\tilde{S}_{T_i} = \frac{(1 - \frac{\text{var}[\mathbb{E}(Y_1|q_{\text{\texttildelow} i})]}{\text{var}(Y_1)}) \times \text{var}(Y_1) + (1 - \frac{\text{var}[\mathbb{E}(Y_2|q_{\text{\texttildelow} i})]}{\text{var}(Y_2)}) \times \text{var}(Y_2)}{ \text{var}(Y_1) +  \text{var}(Y_2)}.  \label{eq: genSobol}
\end{eqnarray}
\noindent
Throughout the rest of this section, we discuss the algorithm presented in \cite{Smith2013} to approximate this formula.

Reconsider a nonlinear input-output relation
\begin{eqnarray}
    Y_j = f_j(q), \quad q = \left[q_1, \dots, q_{11}\right] 
    \label{eq: Sobolnonlineariorelation}
\end{eqnarray}
where $Y_j$ is a scalar response variable and $q_i$ is a model parameter whose index $i$ ($i = 1, \ldots, 11$) corresponds to the indices listed in Table \ref{table: nominal}. We are interested in how the parameters affect both the number of M1 microglial cells and the number of M2 microglial cells, so we integrate the time-dependent response of M1 and M2 microglial cells over the 14 day time period considered in the model to achieve two scalar response variables:
\begin{equation}
   Y_1 = f_1(q) = \int_0^{14} M1(t;q) dt 
   \label{eq: SobolY1}
\end{equation}
and
\begin{equation}   
     Y_2 = f_2(q) = \int_0^{14} M2(t;q) dt .
    \label{eq: SobolY2}
\end{equation}
We allow parameters to vary over a space of $80 -120\%$ their nominal values given in Table \ref{table: nominal}. We further assume that the parameters are each initially scaled to lie in the interval $[0, 1]$ for sampling, then are rescaled to their admissible parameter space before computing the sensitivity measures.

We use MATLAB's \texttt{sobolset} function to generate a quasi-random sample of size $50,000 \times 11$. Half of these samples form the rows of matrix $\mathsf{A}$, which has dimensions $25,000 \times 11$, and the other half form the rows of matrix $\mathsf{B}$, a nonidentical $25,000 \times 11$ matrix. We generate eleven additional matrices, denoted as $\mathsf{C}_1, \ldots, \mathsf{C}_{11}$, such that each $\mathsf{C}_i$ corresponds to the parameter $q_i$ and has its $i^{th}$ row taken as the $i^{th}$ row of $\mathsf{A}$ and its remaining 10 rows taken from $\mathsf{B}$.
We then compute two scalar model outputs for each row of the matrices $\mathsf{A}$, $\mathsf{B}$, and $\mathsf{C}_i$ for all $i$ by first running a forward simulation of the model with parameter values set to the entries in the respective row and then calculating the response variables $Y_1$ and $Y_2$. This results in scalar response vectors of size $1 \times 25,000$ for each matrix, denoted by $Y_{1A}$, $Y_{1B}$, $Y_{1C_i}$, $Y_{2A}$, $Y_{2B}$, and $Y_{2C_i}$, where
\begin{eqnarray}
    Y_{1A} = f_1(\mathsf{A}), \ \ Y_{1B} = f_1(\mathsf{B}), \ \ Y_{1C_i} = f_1(\mathsf{C}_i),
    \label{eq: SobolY1ABC} \\[0.2cm]
    Y_{2A} = f_2(\mathsf{A}), \ \ Y_{2B} = f_2(\mathsf{B}), \ \ Y_{2C_i} = f_2(\mathsf{C}_i). 
    \label{eq: SobolY2ABC}
\end{eqnarray}
Given these response vectors, we approximate $\tilde{S}_{T_i}$ via the following formula:
\begin{equation}
    \tilde{S}_{T_i} =  \frac{S_1 + S_2}{\biggl(\frac{1}{M}Y_{1A}Y_{1A}^T - f_{1_0}^2\biggl) + \biggl(\frac{1}{M}Y_{2A}Y_{2A}^T - f_{2_0}^2\biggl)}
\label{eq: SobolSTi}
\end{equation}
where
\begin{eqnarray}
    S_1 &=& \biggl(1 - \Frac{\frac{1}{M}Y_{1B}Y_{1C_i}^T - f_{1_0}^2}{\frac{1}{M}Y_{1A}Y_{1A}^T - f_{1_0}^2}\biggl) \times \biggl(\frac{1}{M}Y_{1A}Y_{1A}^T - f_{1_0}^2\biggl) \\[0.2cm]
    S_2 &=& \biggl(1 - \Frac{\frac{1}{M}Y_{2B}Y_{2C_i}^T - f_{2_0}^2}{\frac{1}{M}Y_{2A}Y_{2A}^T - f_{2_0}^2}\biggl) \times \biggl(\frac{1}{M}Y_{2A}Y_{2A}^T - f_{2_0}^2\biggl) \\[0.2cm]
    f_{1_0}^2 &=& \biggl(\frac{1}{M} \sum_{k=1}^M Y^{k}_{1A}\biggl) \times \biggl(\frac{1}{M} \sum_{k=1}^M Y^k_{1B}\biggl)  \\[0.2cm]
    f_{2_0}^2 &=& \biggl(\frac{1}{M} \sum_{k=1}^M Y^{k}_{2A}\biggl) \times \biggl(\frac{1}{M} \sum_{k=1}^M Y^k_{2B}\biggl) 
\end{eqnarray}
and $M = 25,000$.

\begin{figure}[t!]
 \centerline{\includegraphics[width=0.8\linewidth]{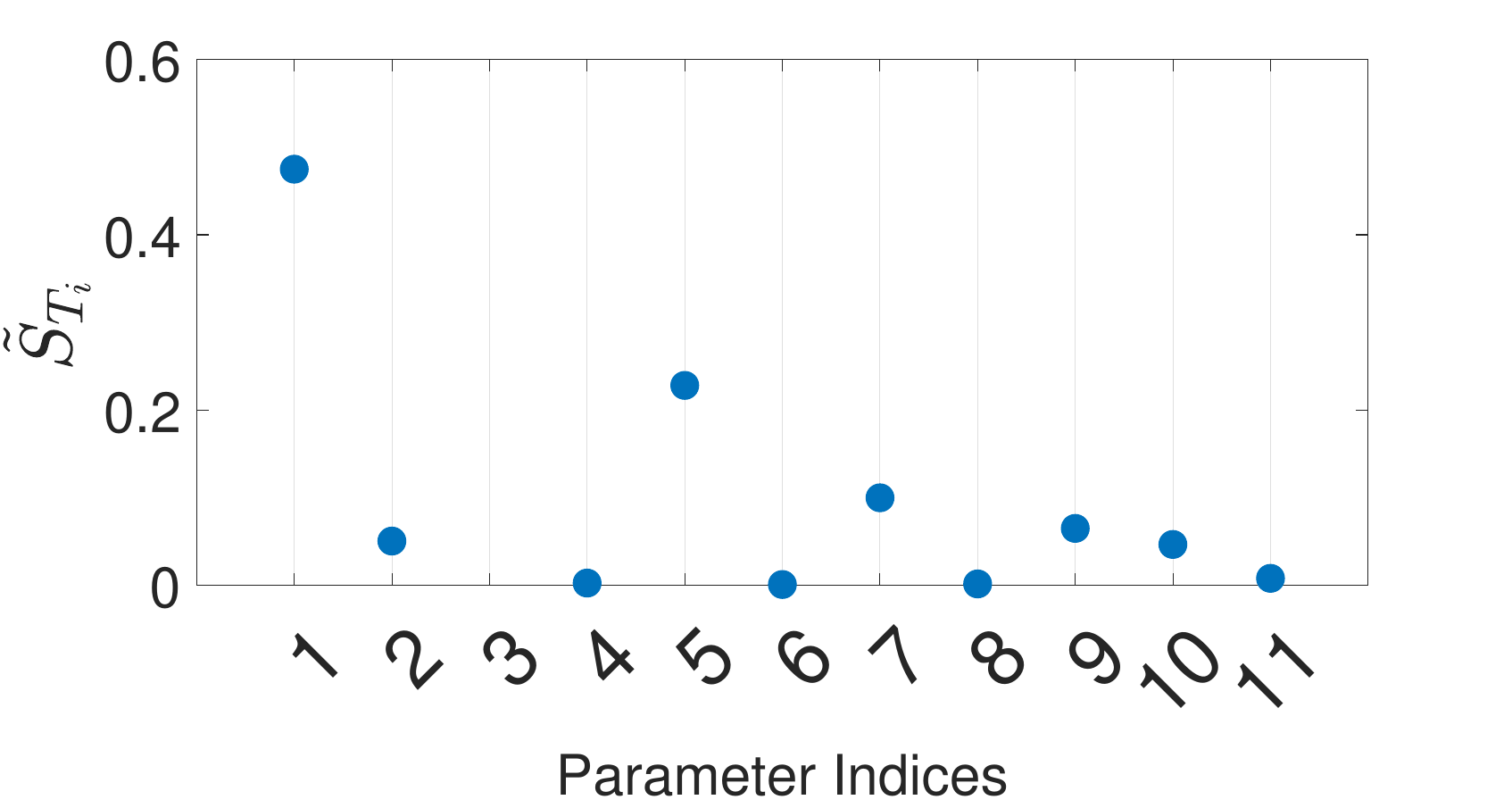}}
	\caption{Results of generalized Sobol sensitivity analysis for each parameter $q_i$ in the model \eqref{eq: modelM1}--\eqref{eq: modelM2} with time-dependent activation functions given in equations \eqref{eq: fM1} and \eqref{eq: fM2} and indices corresponding to those in Table \ref{table: nominal}.}
	\label{fig: GSS}
 \end{figure}

Figure \ref{fig: GSS} shows the results of the generalized Sobol sensitivity analysis. The five most sensitive parameters in order from most sensitive to least sensitive are: $a$, $L_{M2}$, $t_{M2}$, $s_{M2 \rightarrow M1}$, and $L_{M1}$. Recall that parameters $a$, $L_{M1}, L_{M2}$, and $t_{M2}$ all relate to the time-dependent activation of resting microglial cells to the M1 and M2 phenotypes, and parameter $s_{M2 \rightarrow M1}$ is the switching rate from the M2 phenotype to the M1 phenotype.

\subsection{Markov Chain Monte Carlo Estimation Procedure} %

After performing sensitivity analysis, we use Markov Chain Monte Carlo (MCMC) sampling to obtain a probability distribution for the five most sensitive parameters. More specifically, we apply the Metropolis-Hastings (MH) algorithm using the MCMC toolbox for MATLAB \citep{mcmcstat}. In the following section, we summarize the MCMC procedure; however, we encourage interested readers to see, e.g., \cite{Haario2001}, \cite{Haario2006}, and \cite{Robert2004} for more details.

Our goal is to find the set of parameters $\tilde{\theta}$ which minimizes the difference between the model states and the observed data.
Let $\hat{M1}(t_j; \theta)$ and $\hat{M2}(t_j; \theta)$ denote the states of the ODE model in \eqref{eq: modelM1}--\eqref{eq: modelM2} using parameters $\theta$, and let $M1_{obs}(t_j)$ and $M2_{obs}(t_j)$ denote the observed averaged counts of M1 and M2 microglial cells, respectively, as shown in Figure \ref{fig: dataav}.
We assume $t_0 < t_1 < \cdots < t_T$, where $T$ is the time index of the last observation and $j = 1, \ldots, T$.
To fit the proposed ODE model to the averaged experimental data, we aim to find $\tilde{\theta}$ that minimizes the sum of squares cost function
\begin{equation}
ss(\theta) = \sum_{j = 1}^T \Big(M1_{obs}(t_j) - \hat{M1}(t_j; \theta)\Big)^2 + \Big(M2_{obs}(t_j) - \hat{M2}(t_j; \theta)\Big)^2.
\label{eq: MCMCss}
\end{equation}
We note that while the ODE model in \eqref{eq: modelM1}--\eqref{eq: modelM2} depends on the 11 parameters summarized in Table~\ref{table: nominal}, here we focus on estimating only the five most sensitive parameters obtained from the Sobol sensitivity analysis and fix the remaining parameters to their nominal values.

Approaching this problem from a statistical framework, we apply the MH algorithm to approximate $\tilde{\theta}\in\R^5$.
The MH algorithm is a Bayesian sampling method used to obtain a sequence of random samples from the posterior distribution of the parameters $\theta$ given the data $D$. By way of Bayes' Theorem, the posterior distribution $p(\theta|D)$ combines prior beliefs about parameter values $p(\theta)$ with information from the observed data encoded in the likelihood function $p(D|\theta)$. 
To perform MH sampling, we need only know the form of the posterior up to some normalizing constant, so that
\begin{equation}
p(\theta|D) \propto p(D|\theta)p(\theta).
\label{eq: MCMCBayes}
\end{equation}

We prescribe a Gaussian prior distribution for each of the five most sensitive parameters with mean equivalent to the respective nominal value given in Table \ref{table: nominal}. Standard deviations are chosen to obtain distributions which represent parameter values that uphold the model trends discussed in Section \ref{subsec: np} when one of $a$, $L_{M2}$, $t_{M2}$, $s_{M2 \rightarrow M1}$, or $L_{M1}$ varies and all other parameters are set to their nominal values in Table \ref{table: nominal}. Prior distributions for the five most sensitive parameters are summarized in Table \ref{table: prior}.
To determine the likelihood function, we first consider the likelihood of each individual datum. Let $y_j = [M1_{obs}(t_j), M2_{obs}(t_j)]$ and $\hat{x}_j = [\hat{M1}(t_j; \theta), \hat{M2}(t_j; \theta)]$ for $j = 1, \ldots T$. 
Assuming the data are direct observations of the model states with normally-distributed observation error, $y_j = \hat{x}_j + \varepsilon_j$, $\varepsilon_j\sim\mathcal{N}(0,\mathsf{\Sigma})$, the likelihood of each datum, $p(y_j |\theta)$, follows a multivariate normal distribution.
Letting the covariance $\mathsf{\Sigma}$ be equivalent to the $2 \times 2$ identity matrix, it follows that
\begin{align}
    p(D|\theta) &= \prod_{j = 1}^T p(y_j | \theta) \\
    &\propto \prod_{j = 1}^T \exp{\bigg(-\frac{1}{2} (y_j - \hat{x}_j)^T \mathsf{\Sigma}^{-1} (y_j - \hat{x}_j) \bigg)} \\
   &\propto \prod_{j = 1}^T \exp{\bigg(-\frac{1}{2} \bigg[\Big(M1_{obs}(t_j) - \hat{M1}(t_j; \theta)\Big)^2 + \Big(M2_{obs}(t_j) - \hat{M2}(t_j; \theta)\Big)^2\bigg] \bigg)} 
    \\& \propto \exp{\bigg(-\frac{1}{2} \sum_{j=1}^T \Big(M1_{obs}(t_j) - \hat{M1}(t_j; \theta)\Big)^2 + \Big(M2_{obs}(t_j) - \hat{M2}(t_j; \theta)\Big)^2 \bigg)}
    \\& \propto \exp{\bigg(-\frac{1}{2} ss(\theta) \bigg)}.
\end{align}

\begin{table}[t!]
\renewcommand{\arraystretch}{1.25}
\begin{center}
\begin{tabular}{|c c c|} 
 \hline
Index & Parameter & Prior Distribution \\ 
 \hline\hline
1 &  $a$ & $\mathcal{N}(500, 100)$\\ 
 \hline
5 & $L_{M2}$ & $\mathcal{N}(0.75, 0.1)$ \\
\hline
7 & $t_{M2}$ & $\mathcal{N}(3, 0.5)$\\
 \hline
9 & $s_{M2 \rightarrow M1}$ & $\mathcal{N}(0.325, 0.05)$\\
 \hline
2 & $L_{M1}$ & $\mathcal{N}(0.15, 0.03)$\\
 \hline
\end{tabular}
\end{center}
\caption{Prior distributions for the five most sensitive parameters in model \eqref{eq: modelM1}--\eqref{eq: modelM2} with indices corresponding to those in Table \ref{table: nominal}.}
\label{table: prior}
\end{table}

At each step in the MH algorithm, candidate parameters $\theta^*$ are chosen from a proposal distribution $q(\cdot)$. In this work, $q(\cdot)$ is a five-dimensional multivariate Gaussian distribution. We set the mean of the proposal distribution to be equivalent to a vector of the current parameter values and the covariance to be a diagonal matrix with non-zero elements equivalent to each respective parameter's prior variance.
After sampling $\theta^*$, we decide whether to accept the candidate parameters or reject them in favor of the current parameter vector $\theta$ by computing the acceptance probability
\begin{equation}
\alpha = \min\left\{\Frac{p(D|\theta^*)p(\theta^*) q(\theta^*|\theta)}{p(D|\theta)p(\theta)q(\theta|\theta^*)}, 1\right\}
\label{eq: MCMCalpha}
\end{equation}
and generating a uniform random number, $u \in [0,1]$. If $u \leq \alpha$, we accept the candidate parameters by setting $\theta = \theta^*$. If $u > \alpha$, we reject $\theta^*$ and remain at the current point $\theta=\theta$. We proceed in the same fashion for $200,000$ iterations, treating the first $10,000$ sample points as burn-in samples that are disregarded to ensure that our resulting parameter samples come from their respective posterior distributions.

\begin{figure}[t!]
		\centerline{\includegraphics[width=0.33\linewidth]{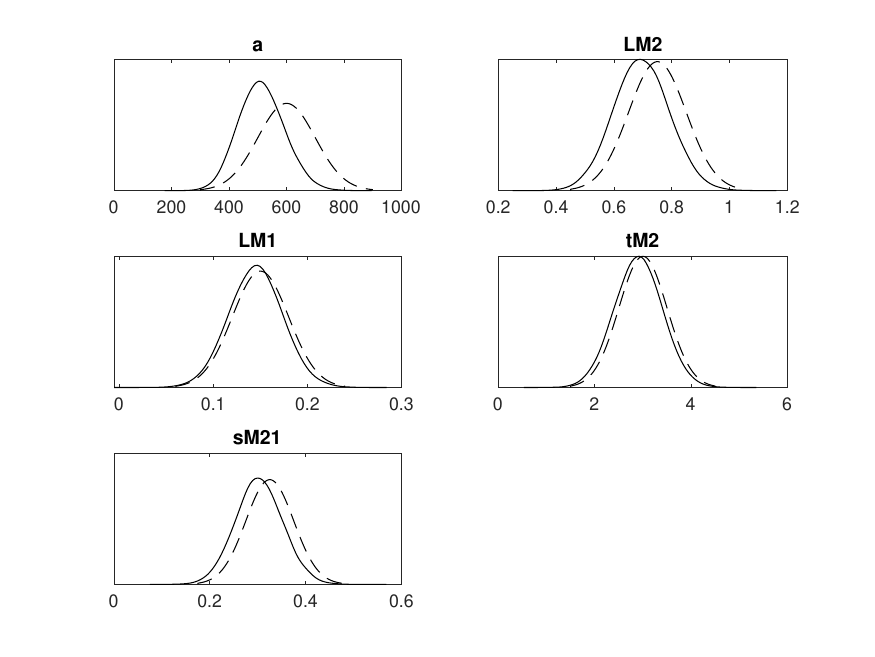} \includegraphics[width=0.33\linewidth]{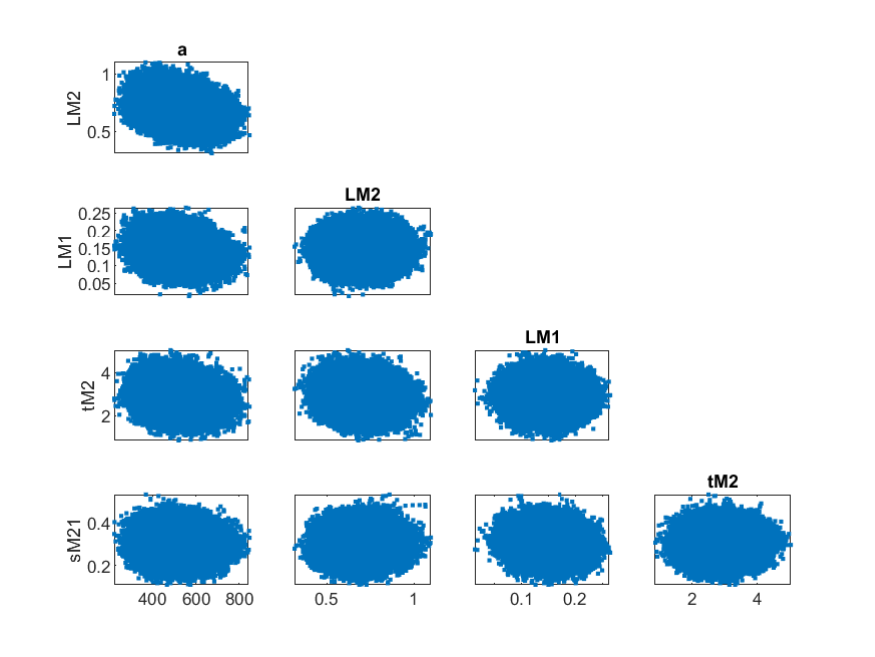} \includegraphics[width=0.33\linewidth]{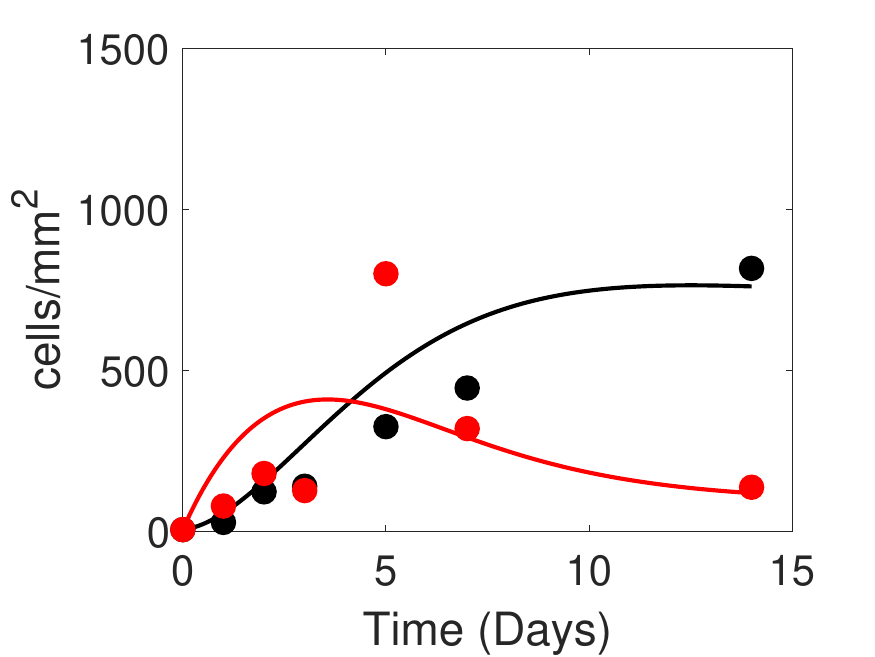}}
		\centerline{\includegraphics[width=0.33\linewidth]{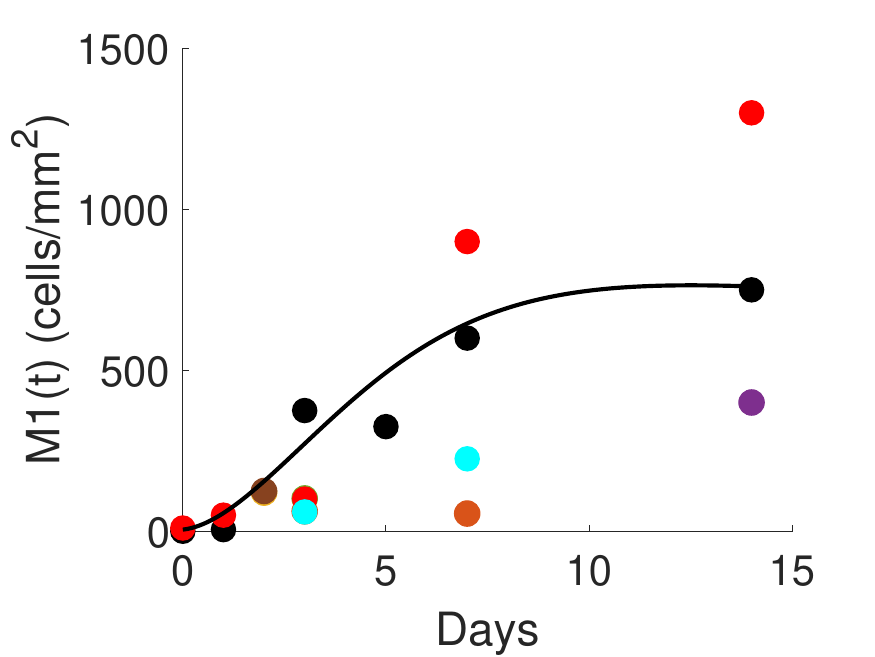} \includegraphics[width=0.33\linewidth]{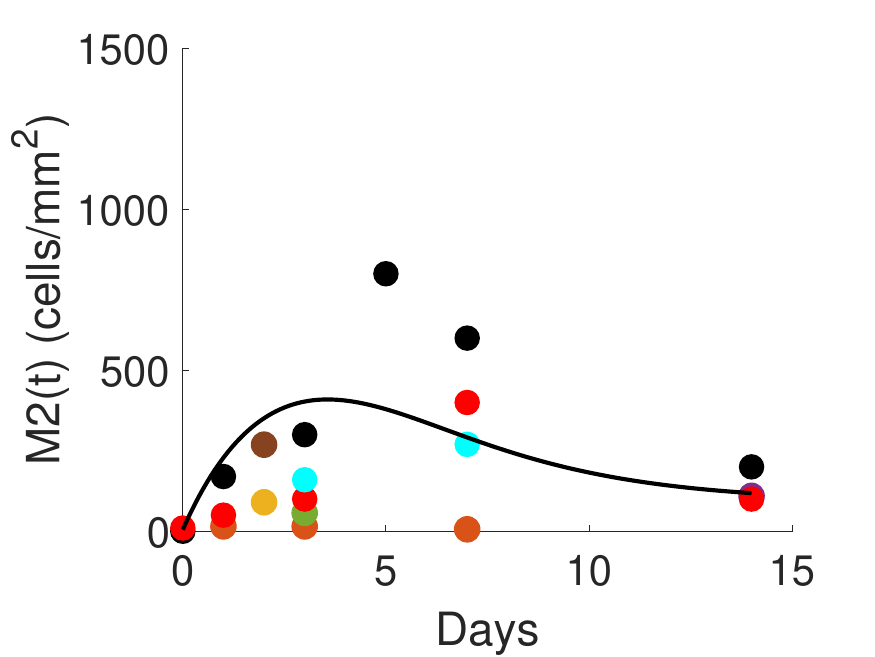}}
		\caption{Results using the Metropolis-Hastings algorithm to estimate five most sensitive parameters with microglial cell measurements at Day 5 considered. On the top row from left to right: the prior (dashed) vs. posterior (solid) distributions, scatter-plots representing the correlation between pairs of parameters, model simulations using nominal parameter values found in Table \ref{table: nominal} and estimated parameters set to the mean of their posterior distribution given in Table \ref{table: mean} plotted with the averaged data. On the bottom row, we plot model simulations with the raw experimental data (colors corresponding to that in Figure \ref{fig: rawdata}).} 
		\label{fig: MH}
	\end{figure}

\begin{figure}[t!]
    \centerline{\includegraphics[width=0.33\linewidth]{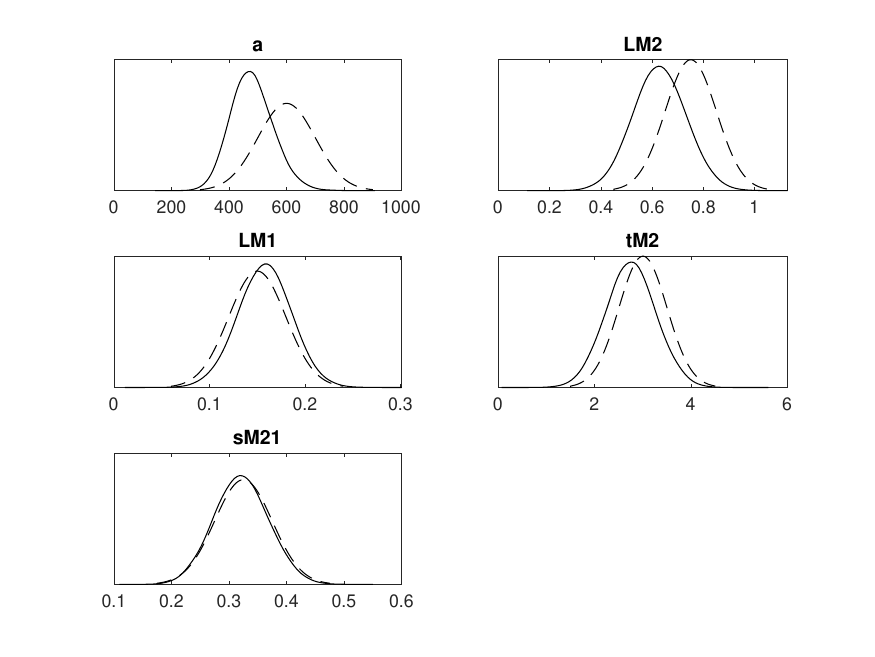} \includegraphics[width=0.33\linewidth]{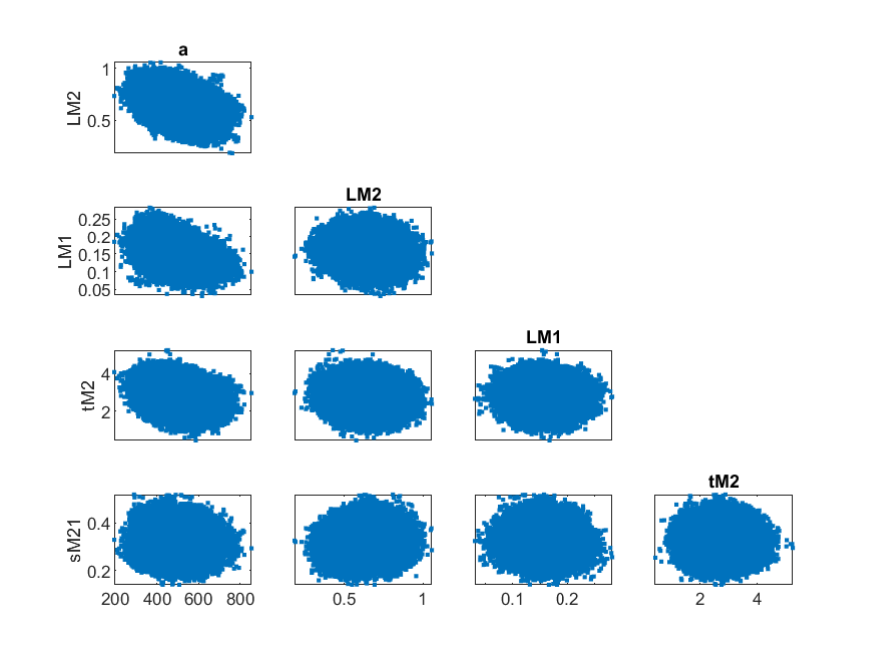} \includegraphics[width=0.33\linewidth]{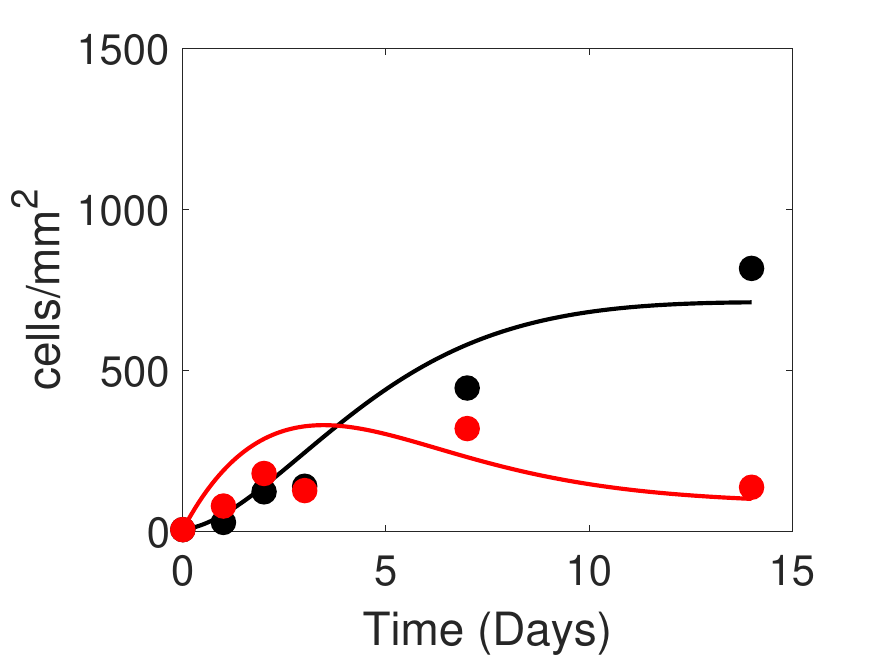}}
	\centerline{\includegraphics[width=0.33\linewidth]{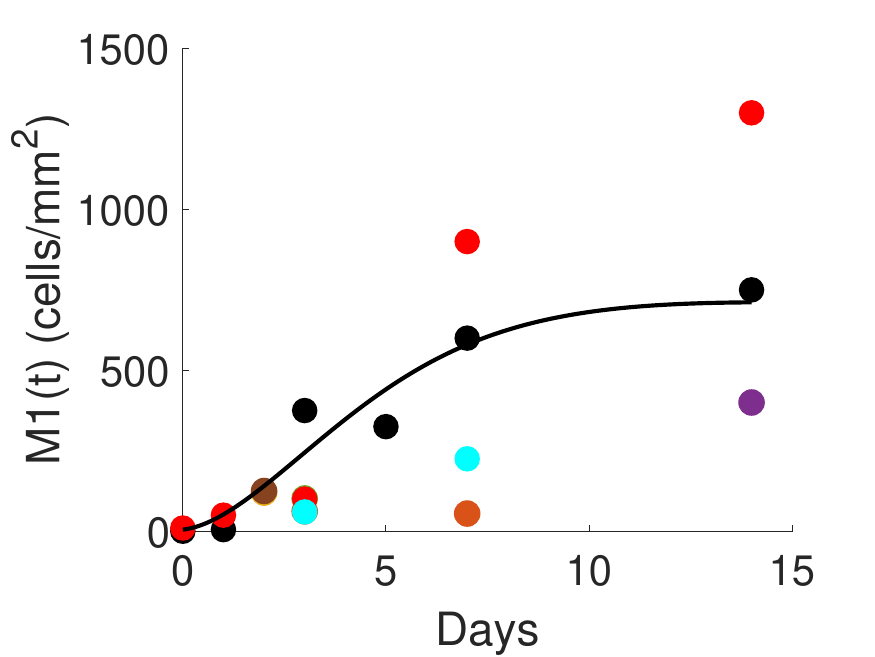} \includegraphics[width=0.33\linewidth]{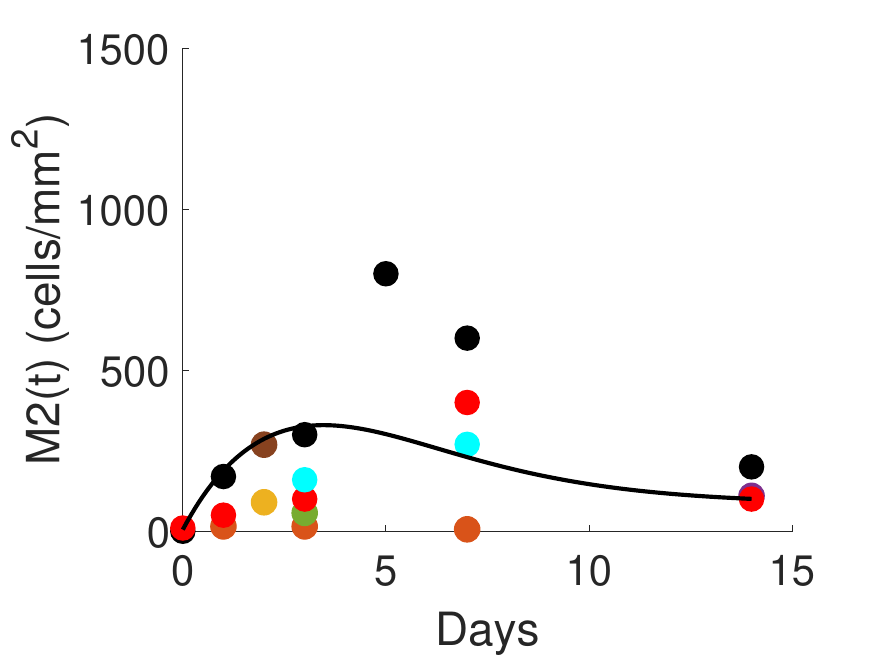}}
	\caption{Results using the Metropolis-Hastings algorithm to estimate five most sensitive parameters while omitting microglial cell measurements at Day 5. On the top row from left to right: the prior (dashed) vs. posterior (solid) distribution, scatter-plots representing the correlation between pairs of parameters, model simulations using nominal parameter values found in Table \ref{table: nominal} and estimated parameters set to the mean of their posterior distribution given in Table \ref{table: mean} plotted with the averaged data. On the bottom row, we plot model simulations with the raw experimental data (colors corresponding to that in Figure~\ref{fig: rawdata}).}
	\label{fig: MHwithout5}
	\end{figure}

Figure \ref{fig: MH} shows the results of estimating the five most sensitive parameters using the MH algorithm with the full set of observed data (including both the number of M1 and M2 microglial cells at Day 5). The histograms show the prior parameter distributions in dashed black lines and the sampled posterior distributions in solid black lines. Note that the posterior distributions for $t_{M2}$ and $L_{M1}$ remain similar to their prior distributions. Additionally, there does not appear to be any strong correlations between pairs of parameters. The resulting model simulations use the mean of the posterior distribution as the point estimate for each of the five estimated parameters (listed in Table \ref{table: mean}), setting all other parameters to their nominal values found in Table \ref{table: nominal} and solving the ODE system in \eqref{eq: modelM1}--\eqref{eq: modelM2} numerically with MATLAB's \texttt{ode15s}. 
The resulting simulations provide a reasonable fit to both the averaged and raw experimental data. In this simulation, M2 is the dominant phenotype for $4.2$ days, which is similar to the model simulation with nominal parameter values showing M2 dominance for $3.9$ days. 
Figure \ref{fig: MHwithout5} shows the output of estimating the five most sensitive parameters using the MH algorithm while omitting both the number of M1 and M2 microglial cells at Day 5, with mean parameter estimates in Table \ref{table: mean}. In this simulation, M2 is the dominant phenotype for the first $3.9$ days before the M1 cells takeover. Note that results are similar when we both omit and include the number of M1 and M2 microglial cells at Day 5.

\begin{table}[t!]
\renewcommand{\arraystretch}{1.2}
\centering
\begin{tabular}{| c | c | c | c | } 
\hline
Index & Parameter & \shortstack{\\ Mean Value Considering\\ Data Point at 5 Days}  & \shortstack{\\ Mean Value Omitting Data \\ Point at 5 Days}\\  
\hline\hline
1 & $a$ & $513.18$ & $479.61$\\
\hline
5 & $L_{M2}$ & $0.70$ & $0.63$\\
\hline
7 & $t_{M2}$ & $2.89$ & $2.75$ \\
\hline
9 & $s_{M2 \rightarrow M1}$ & $0.30$ & $0.32$\\
\hline
2 & $L_{M1}$ & $0.14$ & $0.16$\\
\hline
\end{tabular}
\caption{Posterior means for the five most sensitive parameters used within model simulations shown in Figures \ref{fig: MH} and \ref{fig: MHwithout5}.}
\label{table: mean}
\end{table}

Further note that the cost function in \eqref{eq: MCMCss} is equivalent to $2.6070 \times 10^5$ when we compute the sum of squares error between the numerical solution to model in \eqref{eq: modelM1}--\eqref{eq: modelM2} using the nominal parameter values found in Table~\ref{table: nominal} and the averaged data including data at Day 5. When we omit data at Day 5 in computing the cost function with nominal parameter values, the sum of squares error is equivalent to $8.5453\times 10^4$. When we replace the values of the five-most sensitive parameters with the mean of their respective posterior distribution obtained from MCMC estimation (given in Table~\ref{table: mean}), the sum of squares error reduces to $7.5660\times 10^4$ when data at Day 5 is considered and $1.8326\times 10^4$ when data at Day 5 is omitted.
Moreover, while here we chose nominal parameter values that provide a reasonable initial fit to the experimental data, using different sets of nominal parameter values that provide a less accurate initial fit result in similar MCMC estimates of the most sensitive parameters.

\subsection{Forecast Predictions with Forward Propagation of Uncertainty} 

The MCMC sampling procedure described in the previous section results in a posterior sample for each of the estimated model parameters.
Therefore, there are different combinations of parameter values (in addition to the mean estimates) that can be used from the posterior to simulate the forward model. Additionally, since experimental data are only collected over a two-week period, it is of interest to simulate predicted microglial cell counts beyond the fourteen days of observation. 

\begin{figure}[t!]
\centerline{\includegraphics[width=0.5\linewidth]{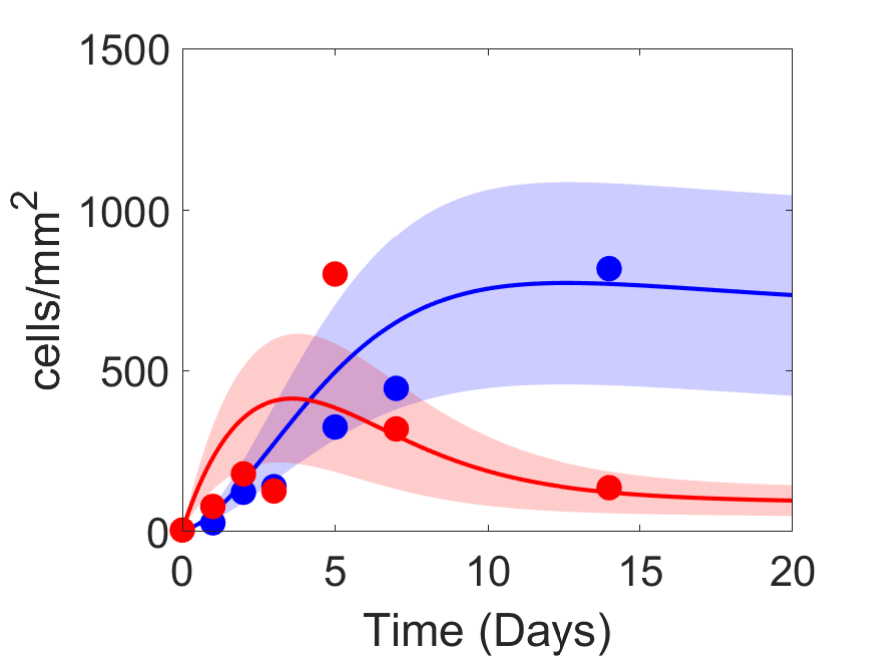} \includegraphics[width=0.5\linewidth]{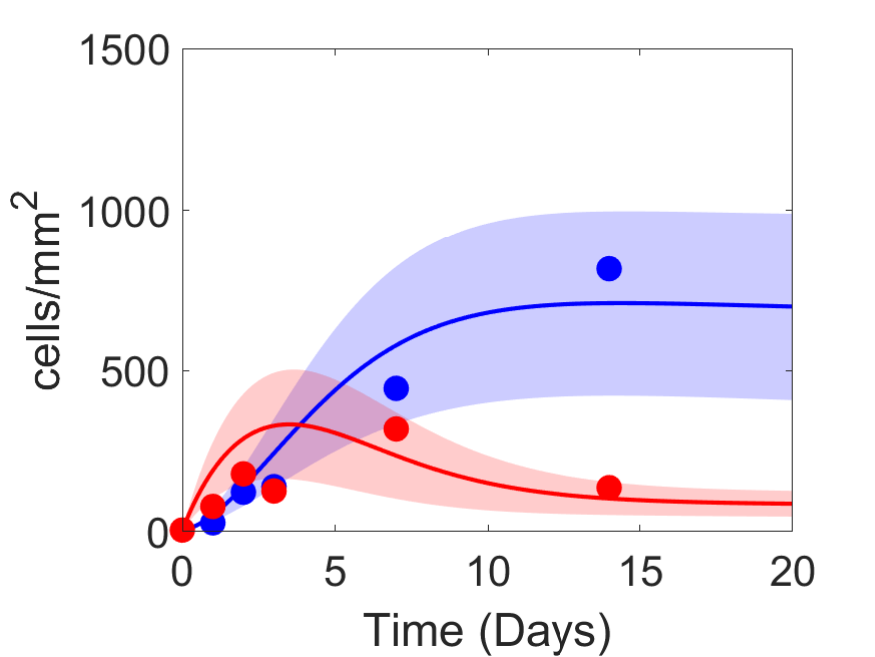}}
\caption{Model predictions with forward propagation of uncertainty when potential outlier data at Day 5 are both included (left) and omitted (right). Blue and red solid lines in each plot show the mean number of M1 and M2 microglial cells, respectively, over a time interval of $[0,20]$ days. The shaded blue and red regions provide uncertainty bounds for M1 and M2 microglial cells, respectively. The blue and red markers show the measured number of M1 and M2 microglial cells given in the averaged data. }\label{fig: predplot}
\end{figure}

In an effort to quantify uncertainty in the model outputs and propagate this uncertainty forward in time with model predictions, we draw $N = 1,000$ random samples from the parameter posteriors shown in Figures \ref{fig: MH} and \ref{fig: MHwithout5}, respectively, and run the forward model using MATLAB's \texttt{ode15s} over the time interval $[0,20]$ days with each of these $N$ parameter sets. After computing the forward model simulations, there are $N$ predicted values that we use to calculate a mean and standard deviation at discretized time points every $0.1$ time units over the interval $[0,20]$. In Figure \ref{fig: predplot}, the blue and red solid lines show the mean estimates of the number of M1 and M2 microglial cells, respectively, at each time point. The blue and red uncertainty bounds are calculated by adding and subtracting two standard deviations of the number of M1 or M2 microglial cells from their respective means. 
We include results when potential outlier data at Day 5 are both included (left) and omitted (right).

The main discrepancies between these two cases are that the uncertainty bounds for M2 are slightly higher between Days 2 and 5 when the data at Day 5 are included and the mean predicted number of M1 cells at Day 14 is closer to the averaged data point at Day 14 when data at Day 5 are included. However, we note that the resulting uncertainty bounds for M2 do not include the potential outlier at Day 5 in either case, suggesting that data at Day 5 may be omitted in future computational studies.

\section{Discussion and Conclusions} \label{sec: Discussion}

Neuroinflammation is a natural response in the penumbra following the onset of ischemic stroke. This process starts with the activation of M1 and M2 microglial cells. In this study, we contribute a compiled set of experimental data on microglial cell counts during MCAO-induced stroke in mice and develop a new, data-inspired mathematical model for microglial cell dynamics in the penumbra based on this data. We derive model equations and set nominal parameter values based on trends observed within the experimental studies. To provide a more robust approach to fit the model parameters and quantify uncertainty in the model predictions, we apply generalized Sobol sensitivity analysis to identify the most influential parameters, estimate the five most influential parameters using MCMC with the averaged experimental data, and simulate forecast predictions with forward propagation of uncertainty.

Simulations using nominal parameter values show that the model captures experimentally-observed trends of M1 and M2 microglial cells. Generalized Sobol sensitivity results indicate that the parameters related to microglial cell activation and switching are the most influential to the outputs of M1 and M2 microglial cells. In particular, the constant source of resting microglial cells ($a$), the maximum value of the activation function for both M1 and M2 ($L_{M1}$, $L_{M2}$), the midpoint of the activation function for M2 ($t_{M2}$), and the rate of M2 to M1 switching ($s_{M2 \rightarrow M1}$) are the most sensitive parameters. Using MCMC to estimate these five influential parameters provides a posterior density of likely values for each parameter. We use the resulting posterior distributions to obtain a level of uncertainty in our numerical solutions and propagate this uncertainty forward in time. 
Considering the experimental data at Day 5 as potential outlier observations, we test both omitting and including this data within our computational framework. Since the results are similar in both cases, we conclude that the data at Day 5 may be omitted in future computational studies.

\begin{figure}[t!]
\centerline{\includegraphics[width=0.5\linewidth]{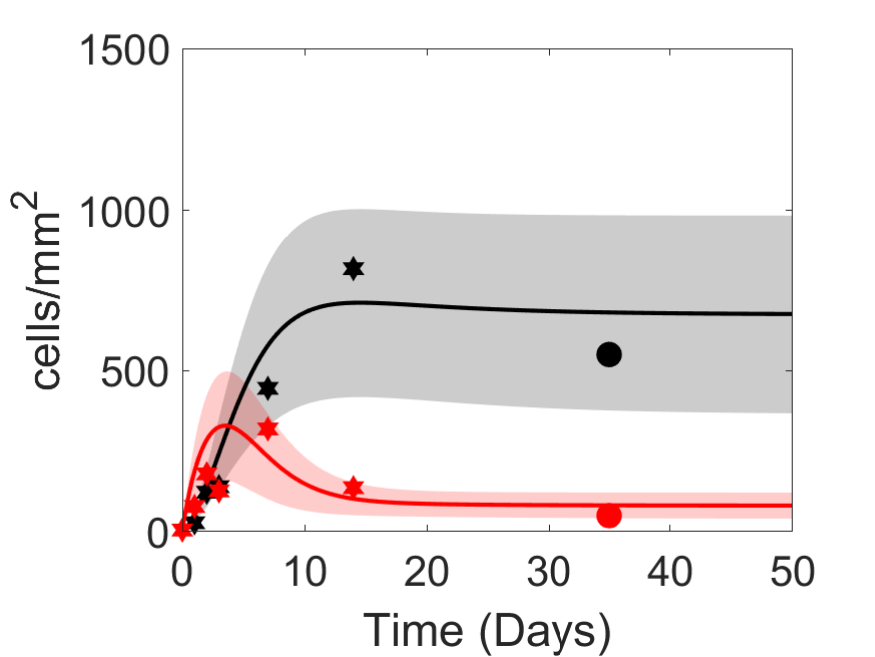}}
\caption{Model predictions with forward propagation of uncertainty when potential outlier data at Day 5 is omitted. Black and red solid lines in each plot show the mean number of M1 and M2 microglial cells, respectively, over a time interval of $[0,50]$ days. The shaded black and red regions provide uncertainty bounds for M1 and M2 microglial cells, respectively. The black and red stars show the measured number of M1 and M2 microglial cells given in the averaged data for Day 0, Day 1, Day 2, Day 3, Day 7, and Day 14. The black and red filled circles at Day 35 are the microglial cell counts given in \cite{Suenaga2015}. }
\label{fig: predplot50}
\end{figure}

To further examine the possible biological implications of the model, Figure~\ref{fig: predplot50} shows forecast predictions with forward propagation of uncertainty over the time interval $[0, 50]$ days, omitting the data at Day 5.
Results emphasize an initial M2 cell dominance of about four days, followed by an eventual takeover of M1 cells. Forecast predictions reflect some trends shown in experimental data papers of ischemic stroke in various brain areas \citep{Suenaga2015, Rupalla1998, Bodhankar2015}. In particular, the forecast uncertainty bounds capture the M1 and M2 microglial cell count at Day 35 reported in \cite{Suenaga2015}. Additionally, model forecasts suggest significantly more M1 cells than M2 cells from fourteen days on and a persistent inflammatory response. Although experimental studies suggest that neuroinflammation may persist past fourteen days, the form of the lingering response is unknown. Our model suggests a near constant number of M1 and M2 microglial cells after the initial two-week period. After Day 35, we may expect the microglial cell numbers to return to baseline values, yet our model remains at steady state with elevated levels of M1 cells.

Due to this behavior, it is not clear that the current model captures biologically-relevant dynamics over longer time intervals. Additional experimental observations are needed for further validation and modification. For example, if experimental studies showed that both M1 and M2 microglial cells return to baseline levels after Day 35, we could adapt the model to account for this by changing the form of time-varying activation function for M1 microglial cells, currently modeled as a logistic function in \eqref{eq: fM1}. Note that the activation function for $f_{M2}(t)$ starts at its maximum value and then returns to zero by the end of the fourteen-day period; however, the activation function for M1 microglial cells starts around zero and then increases to its maximum value before leveling off. This form of activation does not allow the M1 cells to decrease substantially in forecast predictions. Instead, if we modified the form of $f_{M1}(t)$ to be the sum of logistic functions, where the activation again decreases after Day 14, we would be able to capture the data points in \cite{Suenaga2015} as well as model the cell counts returning towards baseline over time.

In future work, we aim to introduce unmeasured neuroinflammatory components into the mathematical model. The two key cellular components considered in this work are M1 and M2 microglial cells; however, other components are produced and interacting with these cells, such as cytokines, macrophages, and neurons, which are not accounted for in the current model. Additionally, there are limitations to the compiled data set and significant differences in the observed cell counts across studies. We aim to further explore experimental studies in the literature and determine a way to account for these discrepancies within the compiled averaged data.


\section*{Acknowledgments}

Many thanks to Dr. Nils Henninger (University of Massachusetts Chan Medical School) for the helpful discussions and clinical perspective on this work.


\section*{ORCID iDs}

Andrea Arnold: \url{https://orcid.org/0000-0003-3003-882X}



\bibliography{paper_refs2.bib}


\end{document}